\documentclass[journal]{IEEEtran}

\usepackage{cite}

\ifCLASSINFOpdf

\else

\fi

\usepackage{amsmath}
\usepackage{amsfonts}
\usepackage{amssymb}
\usepackage{bm}
\usepackage{graphicx}
\usepackage{subcaption}

\usepackage{float}

\usepackage{algorithm}
\usepackage{algorithmicx}
\usepackage{algpseudocode}

\begin{document}
%
\title{Evaluation of  Lane Departure Correction Systems Using a Stochastic Driver Model}
\author{Wenshuo~Wang
	 and Ding~Zhao 
	\thanks{W. Wang is with School of Mechanical Engineering, Beijing Institute of Technology, Beijing, China, 100081, and also with Department of Mechanical Engineering, University of California at Berkeley, Berkeley, CA, US, 94706. e-mail: wwsbit@gmail.com}
	\thanks{D. Zhao (\textit{corresponding author}) is with University of Michigan Transportation Research Institute, Ann Arbor, MI 48109-2150, USA. e-mail: zhaoding@umich.edu}
	}

\markboth{}
{Shell \MakeLowercase{\textit{et al.}}: Bare Demo of IEEEtran.cls for IEEE Journals}

\maketitle

\begin{abstract}
Evaluating the effectiveness and benefits of driver assistance systems is crucial for improving the system performance. In this paper, we propose a novel framework for testing and evaluating lane departure correction systems at a low cost by using lane departure events reproduced from naturalistic driving data. First, 529,096 lane departure events were extracted from the Safety Pilot Model Deployment (SPMD) database collected by the University of Michigan Transportation Research Institute. Second, a stochastic lane departure model consisting of eight random key variables was developed to reduce the dimension of the data description and improve the computational efficiency. As such, we used a bounded Gaussian mixture model (BGM) model to describe drivers' stochastic lane departure behaviors.  Then, a lane departure correction system with an aim point controller was designed, and a batch of lane departure events were reproduced from the learned stochastic driver model. Finally, we assessed the developed evaluation approach by comparing lateral departure areas of vehicles between with and without correction controller. The simulation results show that the proposed method can effectively evaluate lane departure correction systems.
%
%
\end{abstract}


\IEEEpeerreviewmaketitle

\section{Introduction}
\subsection{Motivations}
\IEEEPARstart{S}{ingle-vehicle} road departure accounts for approximately 37.4\% of all fatal vehicle crashes in the United States. Many studies have been conducted on the lane departure warning system, lane keeping assistance system, and lane departure assistance system \cite{son2015robust,merah2016new,cerone2009combined} to warn or assist drivers in keeping the vehicles within the driving lane, preventing the run-off-road crashes. These systems have the potential to address a large proportion of serious injury and fatal crashes. Minoiu Enache \cite{enache2009driver} \textit{et al}. designed a switching steering assistant controller for lane departure cases when drivers have a lapse of attention. Alirezaei\cite{alirezaei2012robust} \textit{et al}. developed a robust controller of steering assistance systems for lane departure avoidance by considering the main uncertainties affecting the vehicle dynamics. Reagan and McCartt \cite{reagan2016observed} undertook an investigation into the frequency of activating a lane departure warning system and a forward collision warning system, concluding that the activation rate is much higher for forward collision warning than lane departure warning. 


\subsection{Related Research}
A well-designed LDC system should be favored by various drivers. However, limited numbers of literature concern on how to evaluate the effectiveness and benefits of these systems. Most research evaluated their LDC systems by conducting real-life experiments. For example, Kwon and Lee \cite{kwon2002performance} evaluated two heuristic decision making strategies--a lateral offset based strategy and a time-to-lane crossing  based strategy--for lane departure warning in real expressway experiments. For setting or computing the time-to-lane crossing, Mammar\cite{mammar2006time}, \textit{et al}. made a systematically theoretical analysis and developed a calibration method by taking many road experiments. Usually, evaluating the LDC systems requires a series of repeated experiments and takes researchers many resources and time, especially in the development stage of a new LDC system. 

Several naturalistic field operational tests (NFOT) have been conducted by the University of Michigan Transportation Research Institute, sponsored by the National Highway Traffic Safety Administration\cite{leblanc2006road1,leblanc2006road2}. However, this type of evaluation method ordinarily has a very high cost such that it is only suitable to test and evaluate the final product, but not affordable to be used in the development procedure. Therefore, approaches that can reuse the NFOT data on newly developed devices will be of great benefits. Recently, the simulation-based evaluation method has become popular \cite{Zhao2016AcceleratedTechniques, Zhao2015AcceleratedData, Zhao2016AcceleratedManeuvers,harding2014vehicle} in the safety of automated vehicles. For example, the vehicle-in-the-loop  method has been developed to investigate driver behaviors and evaluate the ADASs\cite{karl2013driving} by combining real driving experience with the replicability and safety of simulators, which enables the safe and controlled replication of specific traffic situations. Another approach is based on driving simulators \cite{saito2016driver,wanghuman,wang2015study}, in which the driving environment is usually repeatable. Therefore, to make lane departure events as similar to what it might be in the real world, a mathematical model is developed in this paper, capable of regenerating drivers' lane departure behaviors. The lane departure events generated from the developed model can be used to test LDC systems, compare designs,  and evaluate their social benefits. This model may also facilitate research to investigate the physiological and cognitive behaviors of human drivers. 

\subsection{Contributions}
\begin{figure}[t]
	\centering
	\includegraphics[width=\linewidth]{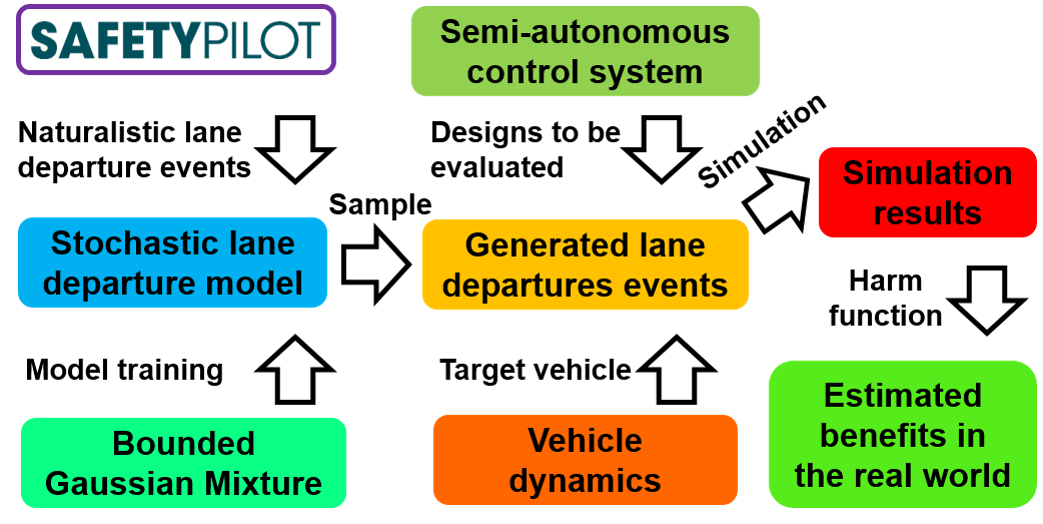}
	\caption{Evaluation procedure using a stochastic lane departure model based on naturalistic driving data.}
	\label{fig:procedure}
\end{figure}

 In this paper, we propose a simulation-based framework to evaluate the LDC systems based 529,096 lane departure events collected from the real world naturalistic driving. The evaluation procedures are shown in Fig. \ref{fig:procedure}. First, a stochastic lane departure model is built based on the bound Gaussian mixture (BGM) model using a very large quantity of  naturalistic driving data. In the trained model, we use a dimension reduction method and apply eight statistical variables representing a lane departure behavior. Thus, from the trained stochastic lane departure model, we extract and regenerate the lane departure event being as similar to what the driver will do in the real world. The lane departure events generated from the stochastic driver model are then used to test and evaluate the LDC systems. When a vehicle is going to drift and cross the lane marker, the LDC systems will automatically control the power steering and bring the vehicle back to the center of the driving lane. In this way, we can repeatedly test and evaluate different LDC systems in various lane departure behaviors. Overall, our contributions are that (1) we propose a new simulation-based evaluation framework to reuse the naturalistic driving data (Fig. \ref{fig:procedure}), (2) a stochastic lane departure model is developed with dimension reduction, and (3) the BGM model is introduced to characterize the statistical feature of the departure behavior.
%
%

\subsection{Paper Organization}
The remainder of this paper is organized as follows. Section II introduces the structure of the stochastic driver model for lane departure behaviors. Section III describes a model fitting approach for the stochastic driver model. Section IV shows the vehicle dynamics model and the designed controller for evaluating the proposed method. Section V discusses and analyzes the experiment results and Section VI gives a further conclusion of this research.

\section{Stochastic Lane Departure Model}
\begin{figure}[t]
	\centering
	\includegraphics[width=\linewidth]{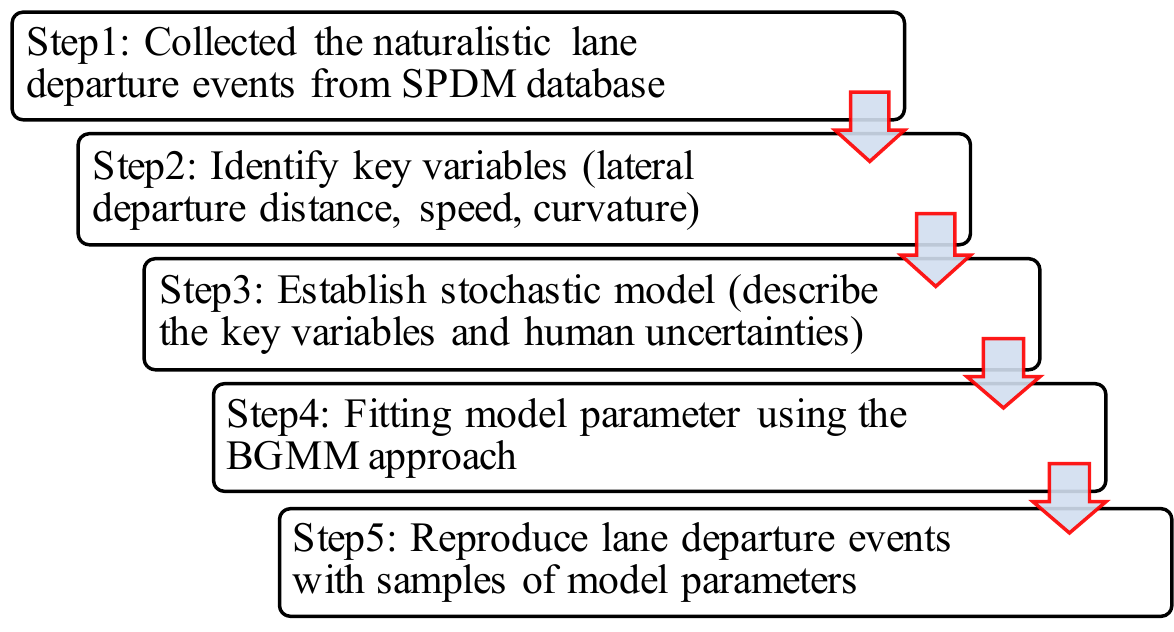}
	\caption{Procedure to build the lane departure model.}
	\label{fig:modelProcedure}
    \end{figure}

The procedure of generating lane departure behavior is shown in Fig. \ref{fig:modelProcedure}.  All the naturalistic driving data were collected from the Safety Pilot Model Deployment (SPDM) database \cite{bezzina2014safety}. In this section, we present approaches for Step 2 and Step 3 in Fig. \ref{fig:modelProcedure}.

\subsection{Key Variables Selection}

    \begin{figure}[t]
	\centering
	\includegraphics[width=1\linewidth]{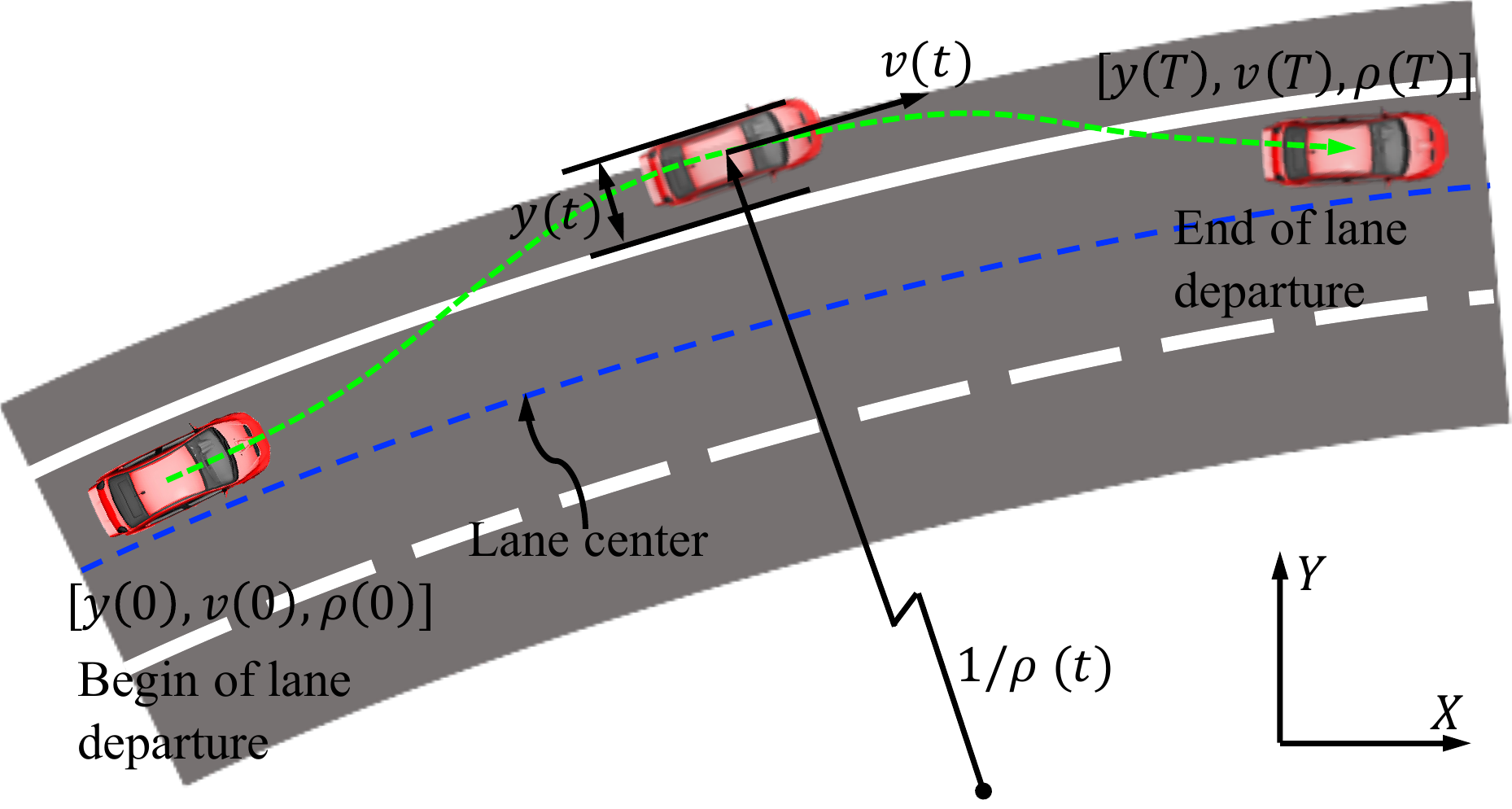}
	\caption{Illustration of a left lane departure event.}
	\label{fig:LD_model}
    \end{figure}

\begin{figure}[t]
	\centering
	\begin{subfigure}{0.48\textwidth}
		\includegraphics[width=\linewidth]{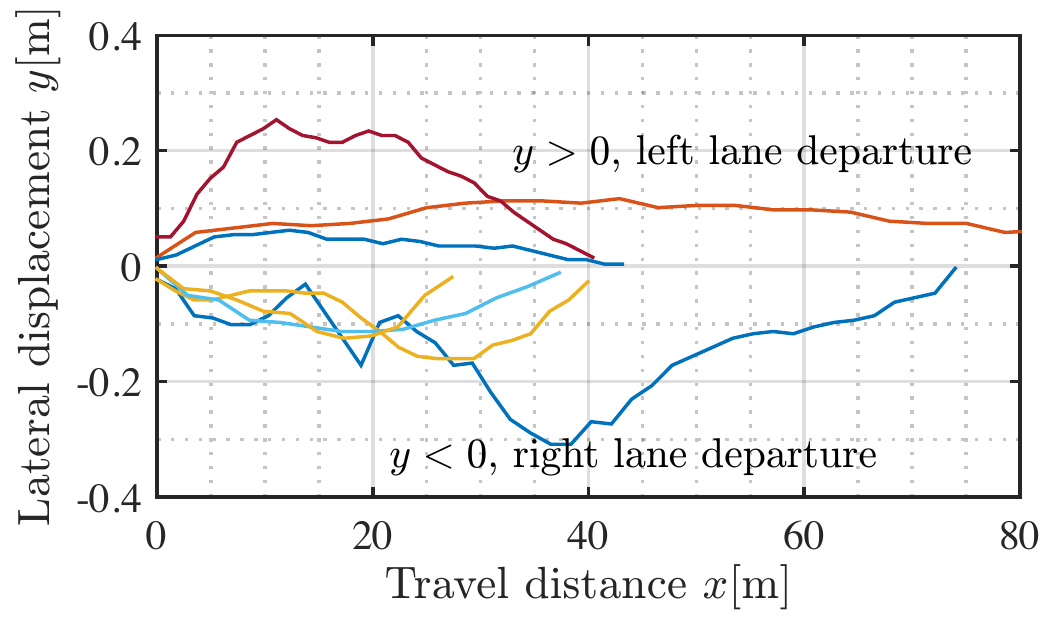}
		\caption{Lateral departure distance}
	\end{subfigure}
	\begin{subfigure}{0.48\textwidth}
		\includegraphics[width=\linewidth]{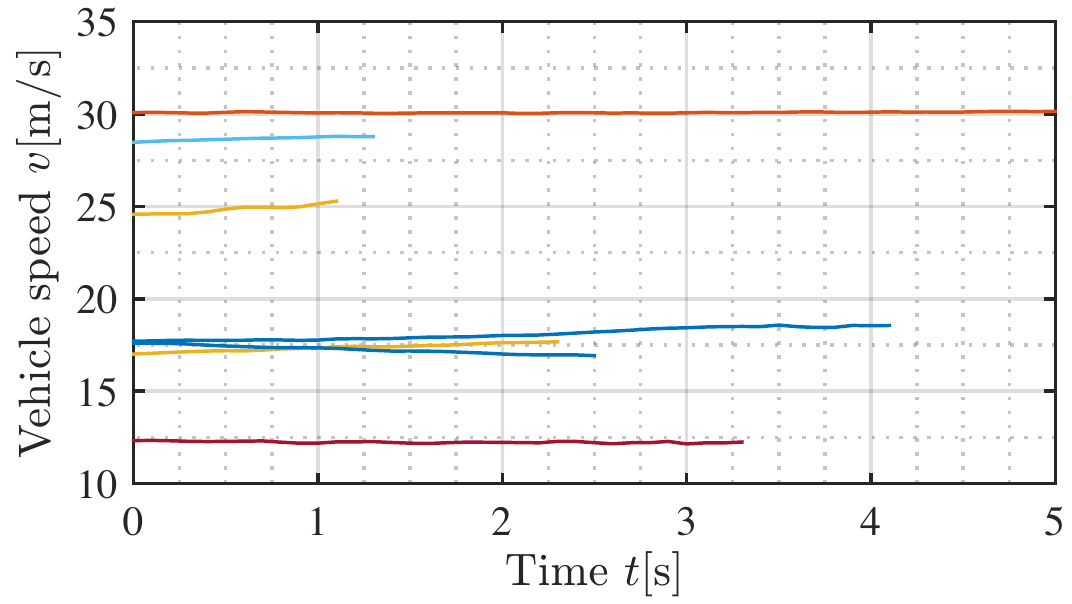}
		\caption{Velocity}
	\end{subfigure}
	\begin{subfigure}{0.48\textwidth}
		\includegraphics[width=\linewidth]{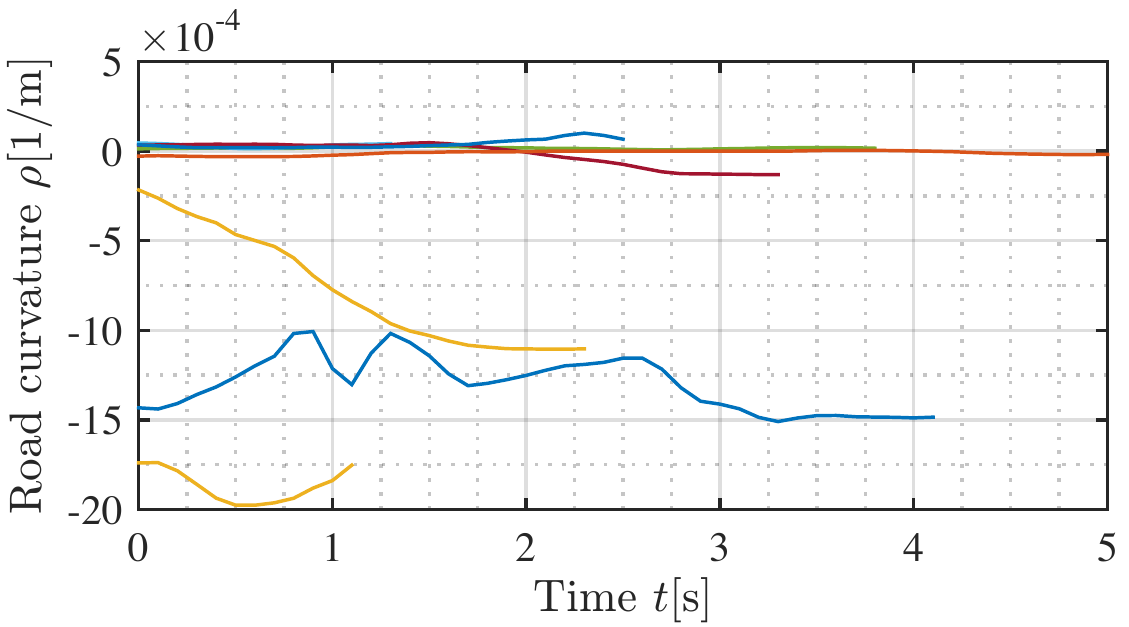}
		\caption{Road curvature}
	\end{subfigure}
	\caption{Lateral offset examples during lane departure.}
	\label{fig:variables}
\end{figure}
%
%
The characteristics of lane departure behavior vary across drivers, even for a single driver with different driving times and locations. For example, due to different driving styles and their diversity in physical and mental states\cite{leblanc2006road1}, the lasting time of a lane departure event differs. Fig. \ref{fig:LD_model} illustrates the left lane departure event, in which three key variables are presented. Fig. \ref{fig:variables} draws a few of left/right lane departure event with three variables, including lateral departure distance $ y $, velocity $ v $, and curvature $ \rho $. In this research, these three key variables (shown in Fig. \ref{fig:LD_model}) are selected to describe a lane departure event and discussed as follows: 

\begin{itemize}
	\item \textit{\textbf{Lateral Departure Distance}} ($y$): Lateral departure distance $ y $ is defined as the distance from the vehicle's left/right side to the left/right lane edge when the vehicle is driving toward left/right. If vehicles are approaching to right lane edge, $ y<0 $, and otherwise, $ y>0 $. 
	\item \textit{\textbf{Vehicle Speed}} ($v$): Vehicle speed has a great influence on the strategy of determining whether the LDC system will be activated. A higher speed normally allows less time to pull the vehicle back to the lane center for drivers.
	\item \textit{\textbf{Lane Curvature}} ($\rho$): According to \cite{mammar2006time}, road curvature is a very important variable to determine and describe a driver's lane departure behavior since the road curvature has a direct impact on the time-to-lane crossing. 
\end{itemize}

\subsection{Dimension Reduction}
Our goal is to build a model that can generate lane departure events being statistically equivalent to the event collected from naturalistic driving data. If the departure duration $T$  is 5 s with sampling time $T_s$ being  0.1 s, we will get $3\times(T/T_s+1)=153$ data points to fully describe the three variables  $y$, $v$, and $\rho$.  This dimension is normally too high and unnecessary to build a stochastic lane departure model. Therefore, it is essential to find a flexible way that can employ all the dataset at a low computational cost. To achieve this, we propose a polynomial fitting-based approach to reduce the data dimension by extracting eight key features of the three variables while reserving the model uncertainty, given by:

\begin{itemize}
	\item \textit{\textbf{Lateral Departure Distance:}} As illustrated in Fig. \ref{fig:variables}(a), $y$  can be approximated as a second order polynomial function of the longitudinal travel distance $x$ with the addition of an error term.
	
	\begin{equation}\label{eq:laterdis}
	y(t)=\tilde{y}(t)+\epsilon_y(t)
	\end{equation}
	\begin{equation}
	\tilde{y}(t)=-\frac{4d_y}{d_x^2}\Big(x(t)-\frac{d_x}{2}\Big)^2+d_y
	\end{equation}
	where $d_x$ is the longitudinal travel distance during the departure, $d_y$ represents the lateral departure calculated by the least square method, and $ \epsilon_y(t) $ is the error term.
    The variance of human driving is captured by the standard deviation of  $\epsilon_y(l)$, calculated from
    
    \begin{equation}
    \label{eq:sigma_y}
    \sigma_y = \sqrt{\dfrac{1}{L-1}\sum_{l=1}^{L}|\epsilon_y(l)-\bar{\epsilon}_y|^2}
    \end{equation}
    where $ L $ is the number of samples in one lane departure event, and  $\epsilon_y(l)=y(l)-\tilde{y}(l)$ represents the error of the $l^{th}$ sample of the lane departure event. In the following paper, we use $t$ to represent continuous time starting from 0 to $ T $ and use $l$ as the index of the discrete sample time starting from 1 to $ L $, and  $\bar{\epsilon}_y=\sum_{l=1}^{L}\epsilon_y(l)$.
     
    \item \textit{\textbf{Velocity:}} From Fig. \ref{fig:variables}(b), we know that the vehicle speed only changes slightly in a lane departure event. Without considering emergency braking behaviors, as such the velocity in a lane departure event can be approximated as a linear function of time, represented by
    
    \begin{equation}
    v(t)=\tilde{v}(t)+\epsilon_v(t)=\bar{a}(t-T/2)+\bar{v} +\epsilon_v(t)
    \end{equation}
    where $\bar{v}=d_x/T$  is the average speed. The average acceleration $\bar{a}$ is estimated using the least square method.
    Similarly, human uncertainties in velocity can be calculated from
     
    \begin{equation}
    \label{eq:sigma_v}
    \sigma_v = \sqrt{\dfrac{1}{L-1}\sum_{l=0}^{L}|\epsilon_v(l)-\bar{\epsilon}_v|^2}
    \end{equation}

    \item \textit{\textbf{Curvature:}} Similar to the vehicle speed, the lane curvature in a departure event only changes slightly and can also be modeled as a linear function of time:
    
\begin{equation}\label{eq:curvature}
\rho(t) = \tilde{\rho}(t)+\epsilon_{\rho}=\frac{\Delta \rho}{T}t+\rho_{0}+\epsilon_{\rho}
\end{equation}
where $\rho_{0}$ is the initial curvature, $\Delta \rho$ is the curvature change from the start to the end of a departure event. To smooth the curvature data, we use linear regression to estimate  $\rho_{0}$ and $\Delta \rho$, such that $\sum_{l=1}^L |\rho(l)-\tilde{\rho}(l)|^2 $ is minimized.
\end{itemize}

\begin{figure}[t]
	\centering
	\includegraphics[width = 0.48\textwidth]{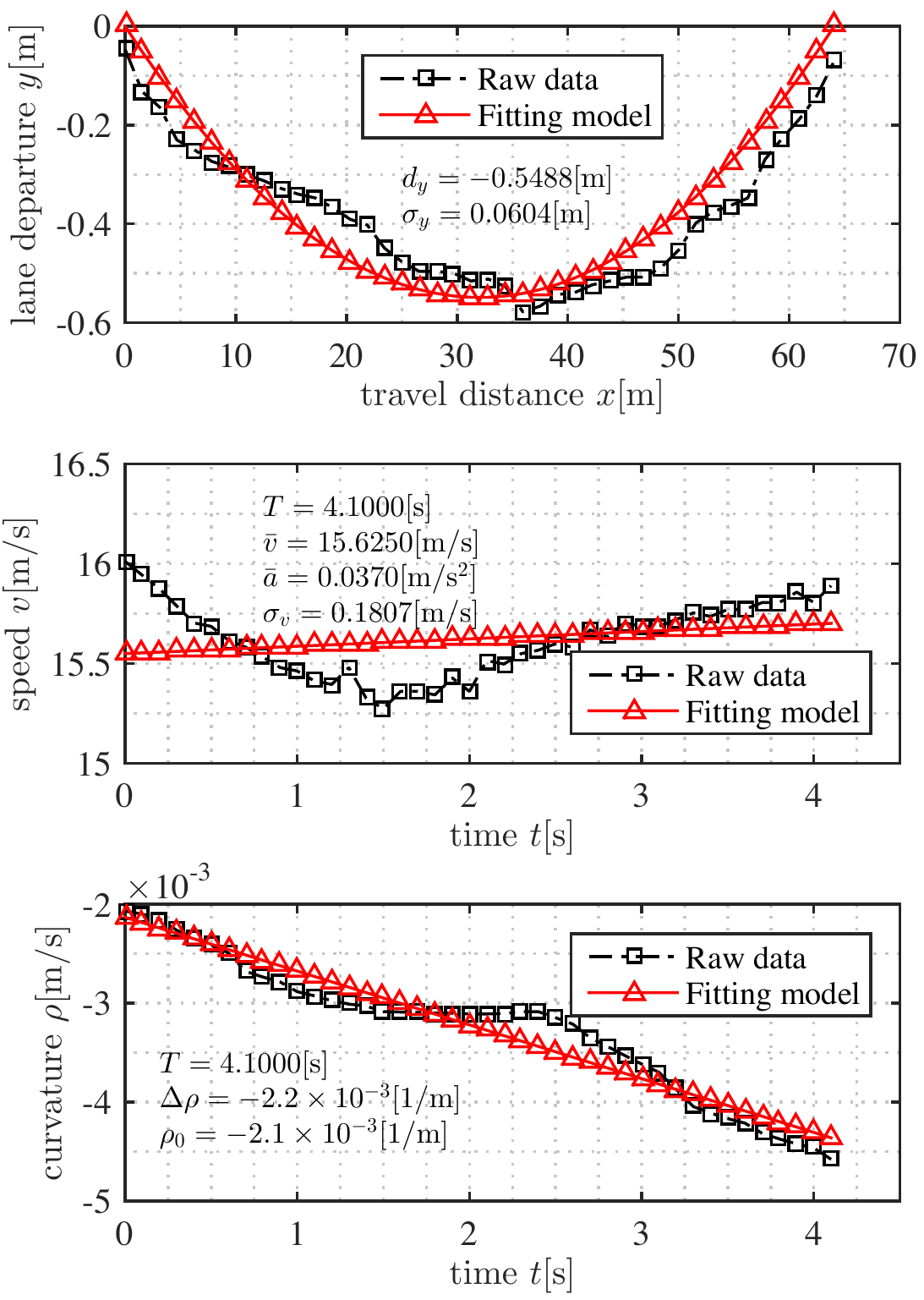}
	\caption{Example of fitting results and identified parameters for a right lane departure event by using our proposed method.}
	\label{fig:Fitting_dimension}
\end{figure}

The advantages of this model with dimension reduced are that (1) it reduces the dimensions of the original data while capturing the stochastic variance of human driving behaviors and (2) each of its parameters has a clearly defined physical meaning. 

Fig. \ref{fig:Fitting_dimension} provides an example of fitting results by using the dimension reduction approach we proposed. We note that for the case with $ T = 4.1 $ s in Fig. \ref{fig:Fitting_dimension}, the approximated approach can capture the lane departure features by using only 8-dimension features (i.e., $ T $, $ d_{y} $, $ \sigma_{y} $, $ \bar{v} $, $ \bar{a} $, $ \sigma_{v} $, $ \Delta \rho $, and $ \rho_{0} $), instead of using 126-dimension features ($ 3\times (T/T_{s}+1) = 126$). Therefore, the proposed dimension reduction approach greatly improves the computational efficiency with reserving the uncertainties of human driving as well as catching the key features of the three variables, i.e., $ y $, $ v $, and $ \rho $. 

    \begin{figure}[t]
    	\centering
    	\includegraphics[width = 0.48\textwidth]{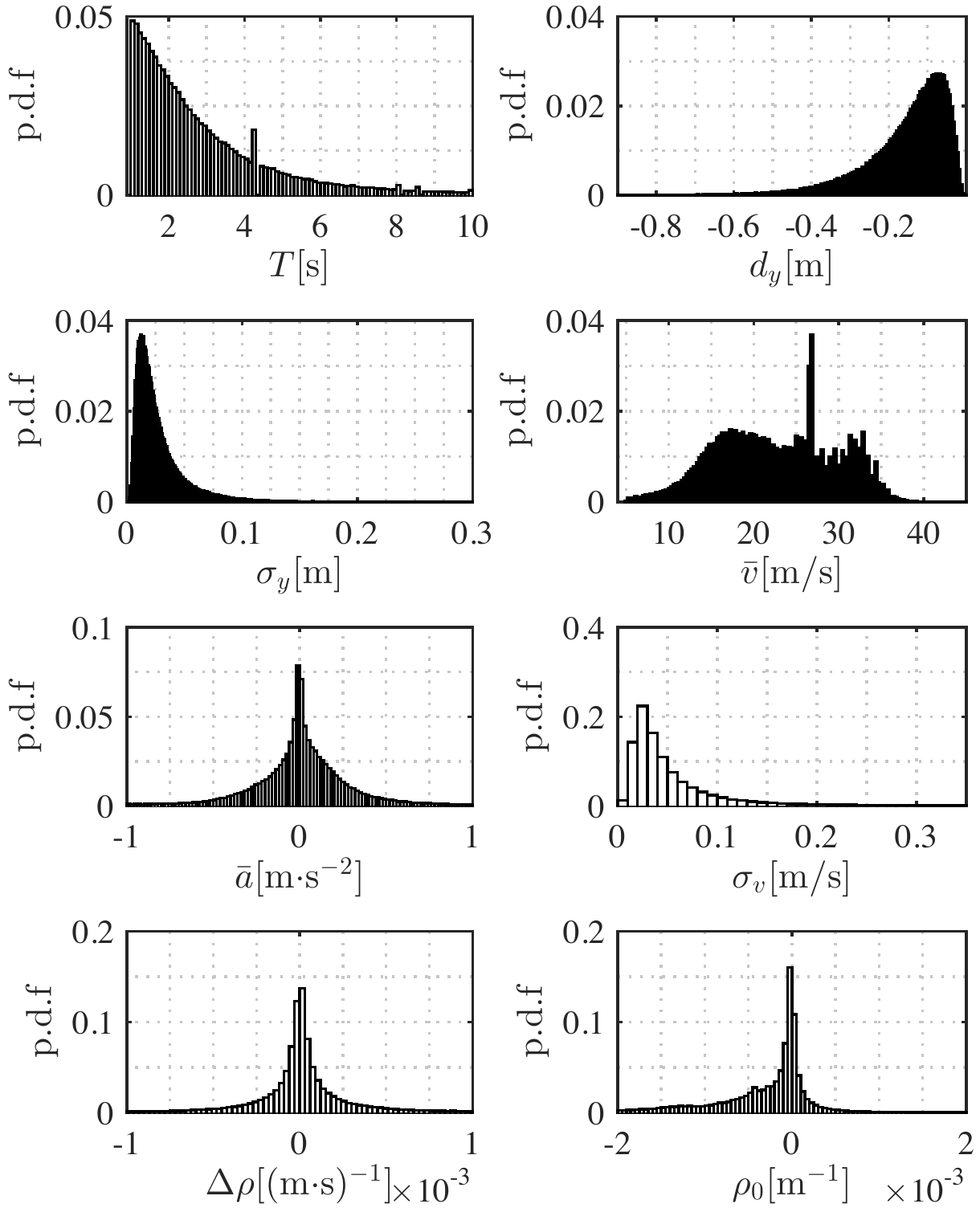}
    	\caption{Marginal distributions of right lane departure variables.}
    	\label{fig:marginal_L}
    \end{figure}
    
    \begin{figure}[t]
    	\centering
    	\includegraphics[width = 0.48\textwidth]{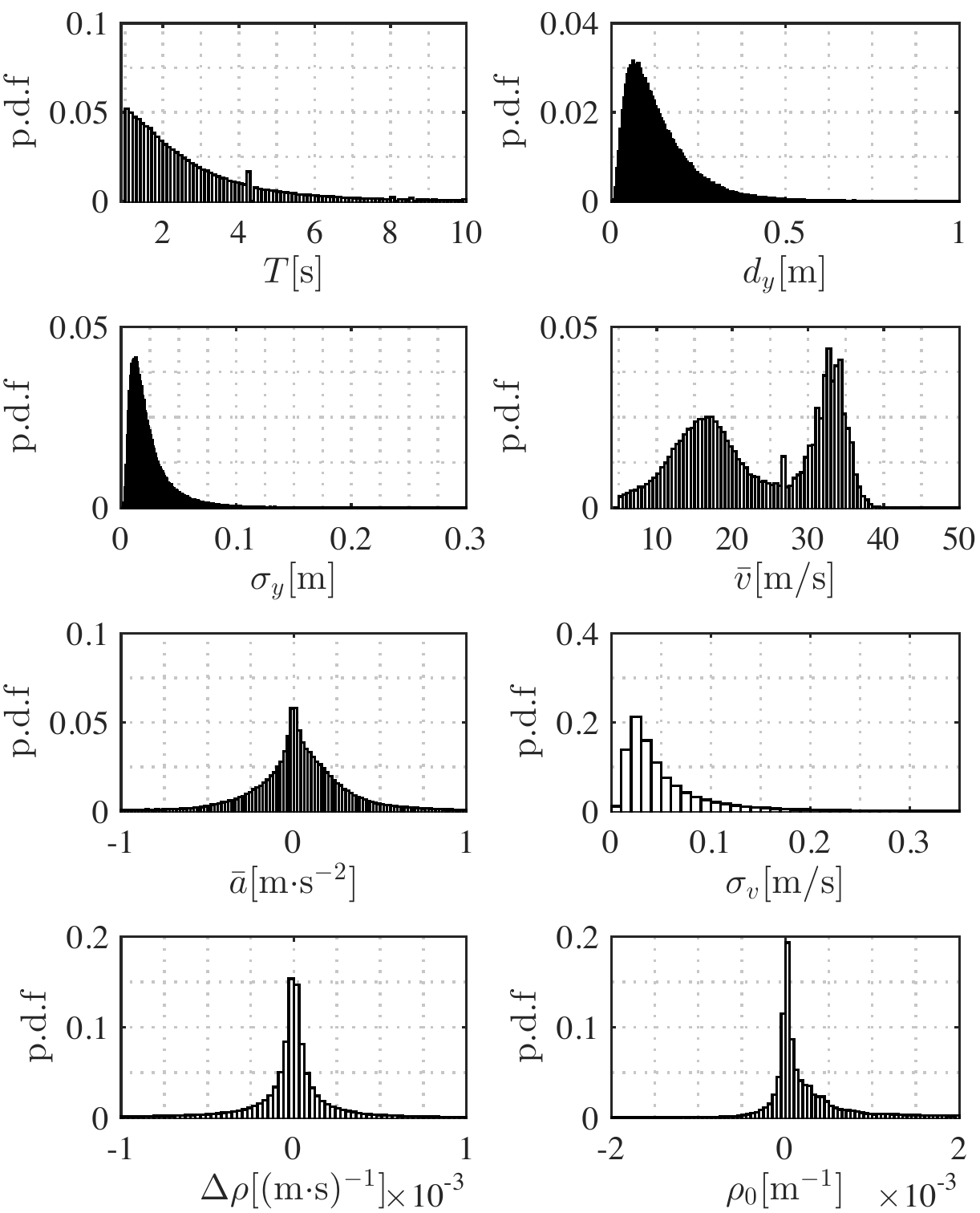}
    	\caption{Marginal distributions of left lane departure variables.}
    	\label{fig:marginal_R}
    \end{figure}

\section{Learning with Naturalistic Driving Data}
In this section, we use the BGM model to capture the stochastic features of these 8 variables mentioned in Section II. And then we can stochastically extract and sample the 8 variables from the trained BGM model, thereby generating various lane departure events.

\subsection{Naturalistic Lane Departure Events}
The naturalistic driving data used in this research are extracted from the SPMD database. It recorded naturalistic driving of 2,842 equipped vehicles in Ann Arbor, Michigan for more than two years. As of April 2016, 34.9 million miles were logged. We used 98 sedans to run experiments and collect the real on-road data. The vehicle is equipped with data acquisition systems and Mobileye$^\circledR$ \cite{harding2014vehicle}. The Mobileye gets the driving data such as lane curvature, lateral displacement with respect to lane marks, and lane tracking measures about the lane delineation both from the painted boundary lines and the road edge, etc. The global navigation satellite system (GNSS) gets the global position (latitude and longitude) and the GNSS time. The CAN-Bus signals including the vehicle speed, acceleration, throttle opening, braking force, and engine speed are obtained.
To ensure consistency of the used dataset, we extract data from the database by the following criteria: 
\begin{enumerate}
	\item The duration of each event should be in the range of 0.5 s to 10 s;
	\item The average velocity of each event should be larger than 5 m/s. This limitation excludes the traffic jam and the stop-and-go behaviors.
\end{enumerate}
In total, 529,096 lane departure events (249,798 left and 279,298 right lane departure events) are identified from 118 drivers over the last four years.

\subsection{Variable Fitting Approach}
\begin{figure}[t]
	\centering
	\includegraphics[width=0.98\linewidth]{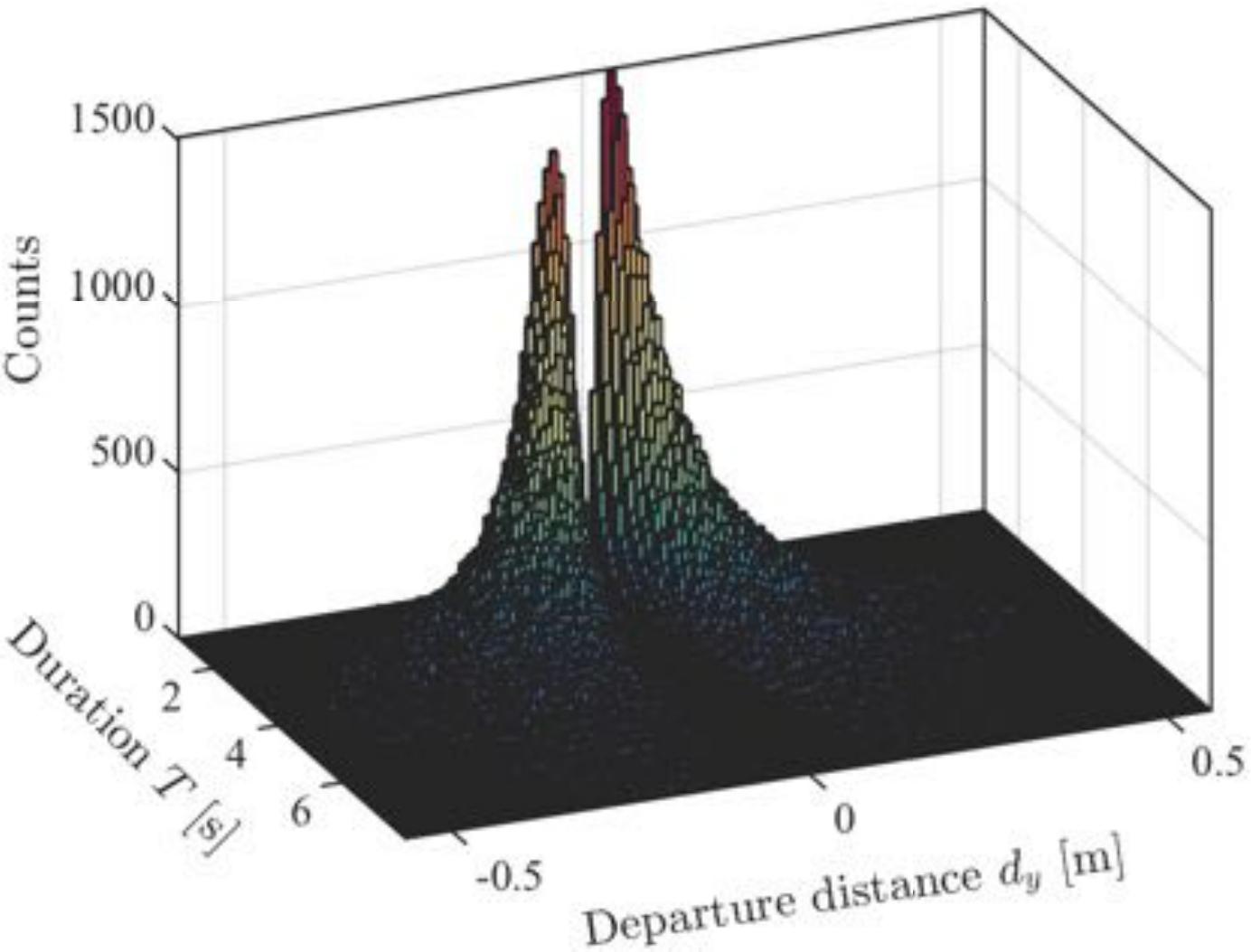}
	\caption{Joint distributions of duration and maximum departure distance for both left and right lane departure cases.}
	\label{fig:joint}
\end{figure}

We aim to develop a statistical model to describe the joint distribution of the eight parameters. Let $\bm{\xi} ^{(n)}=[T^{(n)}$, $d_y^{(n)}$, $\sigma_y^{(n)}$, $\bar{v}^{(n)}$, $\bar{a}^{(n)}$, $\sigma_v^{(n)}$, $\rho_{0}^{(n)}$, $\Delta \rho^{(n)}]^{\top} \in \mathbb{R}^{8\times1} $, where $ \bm{\xi} ^{(n)} $ describes the $ n $th lane departure event and $ n=1$, $2$, $...$, $N $ is the index of departure events. The marginal distribution of $\bm{\xi}$ for left lane departure events and right lane departure events are shown Fig. \ref{fig:marginal_L} and Fig. \ref{fig:marginal_R}, respectively. We can see that each variable follows an unique distribution, especially for the average of vehicle speed $  \bar{v}$ and the initial curvature of departure $ \rho_{0} $. From the joint distribution between duration $ T $ and departure distance $ d_{y} $ as shown in Fig. \ref{fig:joint}, we can see that there is clear dependence between variables. In addition, from Fig. \ref{fig:marginal_L} and Fig. \ref{fig:marginal_R} it can be seen that the distributions of variables $ T $, $ \sigma_{y} $, and $ \sigma_{v} $ have bounded features. Therefore, a flexible probability density function (p.d.f.) is required to model the multi-variate distribution. In this paper, we select and discuss the BGM model to describe the joint distribution of the 8 variables as follows. 

\subsubsection{Structure of the BGM model}
In light of its flexibility and ease of training, the Gaussian Mixture Model (GMM) have been widely used in many applications such as speech recognition \cite{jonas2003bounded}, pattern recognition \cite{nguyen2014bounded}, and driving behaviors \cite{Zhao2015AcceleratedData,wang2017learning}. In terms of driver behavior, some feature boundaries usually exist because of the physical limitations of variables, which thereby makes traditional GMM approaches be difficult to perfectly fit these boundaries. For example, the duration of a departure event should be larger than 0.5 s and less than 10 s, and the vehicle speed should be positive, as shown in Fig. \ref{fig:marginal_L} and Fig. \ref{fig:marginal_R}. To deal with this issue, in this research, we applied the BGM model-based method to incorporating the physical boundaries. One of key advantages of the BGM model is that it considers the variable boundaries while also preserving a tractable form when using the expectation-maximization (EM) algorithm to train the model.

The probability density function of a BGM model can be expressed as \cite{jonas2003bounded}

\begin{equation}
f_{\mathcal{BM}}(\bm{\xi})=\frac{f_{\mathcal{GM}}(\bm{\xi})}{\int_{\bm{b_l}}^{\bm{b_u}}f_{\mathcal{GM}}(\bm{u})d\bm{u}}
\end{equation}
where $ f_{\mathcal{GM}}(\bm{\xi}) $ is a normal GMM, given by

\begin{equation}
f_{\mathcal{GM}}(\bm{\xi}|\bm{\Theta})=\sum_{k=1}^{K}\pi_k g_k(\bm{\xi};\bm{\theta}_k)
\end{equation}
where $ \pi_k \in [0,1] $ are mixing weights with $ \sum_k \pi_k = 1 $, $ g_k $ is the $ k^{\textrm{th}}$ $ D $-dimensional Gaussian distribution component  (In this paper, $ D = 8 $) parameterized by $ \bm{\theta_k}=[\bm{\mu_k},\bm{\Sigma_k}] $, $ \bm{\Theta}=[\pi_1,...,\pi_K,\bm{\theta_1},...,\bm{\theta_K}] $. Here we assume that the boundary is a hyper-rectangle in $ \mathbb{R}^{D\times1} $ with two vertices $ \bm{b_u} =[T^u,d_{y}^{u}, \sigma^{u}, \bar{v}^{u}, \bar{a}^u, \sigma_{v}^{u}, \rho_{0}^{u}, \Delta \rho^{u} ]^{\top} $ and $ \bm{b_l} =[T^{l},d_{y}^{l}, \sigma^{l}, \bar{v}^{l}, \bar{a}^{l}, \sigma_{v}^{l}, \rho_{0}^{l}, \Delta \rho^{l} ]^{\top} $ on the diagonal opposites

\begin{equation}\label{eq:boundary}
\bm{b_l}<\bm{\xi}^{(n)}<\bm{b_u}
\end{equation}
and computed by 

\begin{equation}\label{eq:boundary1}
\begin{split}
{\star}^{u} & = \max \{\star^{(n)},  n = 1,2,\cdots,N\}\\ 
{\star}^{l} & = \min \{\star^{(n)}, n = 1,2,\cdots,N\}\\ 
\end{split}
\end{equation}
where $ \star \in [ T, d_{y}, \sigma_{y}, \bar{v}, \bar{a}, \sigma_{v}, \rho_{0}, \Delta \rho ] $. By incorporating the boundary function, it can be derived that $ f_{\mathcal{BM}} $ is also a mixture

\begin{equation}
f_{\mathcal{BM}}=\sum_{k=1}^{K}\eta_k f_k(\bm{\xi})
\end{equation}
with mixing weights $ \eta_k $ and component density functions $ f_k $:

\begin{align}
\eta_k&=\pi_k\frac{\int_{\bm{b_l}}^{\bm{b_u}}g_k(\bm{u})d\bm{u}}
{\int_{\bm{b_l}}^{\bm{b_u}}f_{\mathcal{GM}}(\bm{u})d\bm{u}}\\
f_k(\bm{\xi})&=\frac{g_k(\bm{\xi})}
{\int_{\bm{b_l}}^{\bm{b_u}}f_{\mathcal{GM}}(\bm{u})d\bm{u}}
\end{align}

\begin{figure}[t]
	\centering
	\includegraphics[width = 0.48\textwidth]{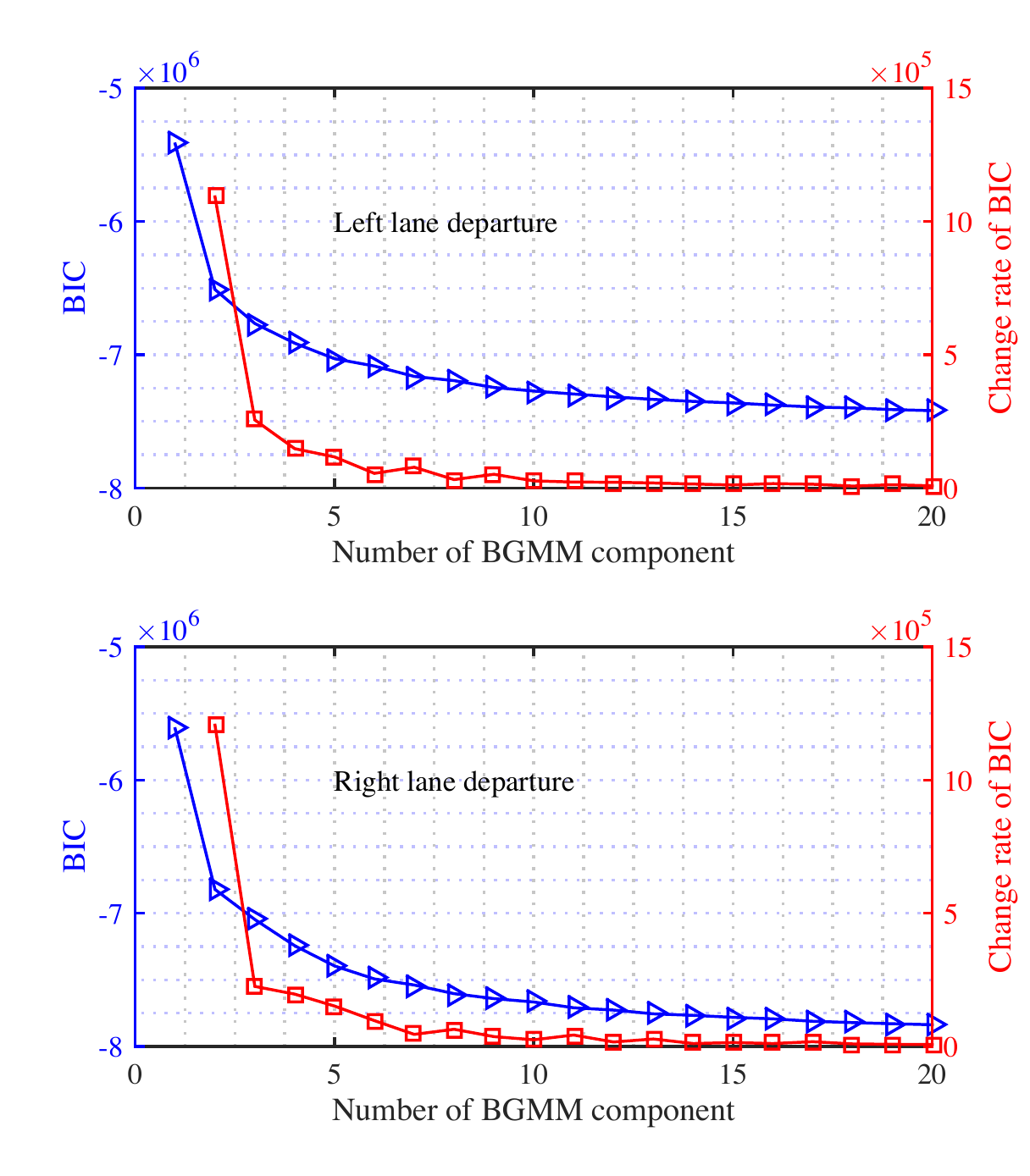}
	\caption{BIC with different numbers of BGM model components for left and right lane departure data.}
	\label{fig:BIC}
\end{figure}

\subsubsection{Parameter Estimation for BGM Model}
The best parameter $ \bm{\Theta}^{\ast} $ for the BGM model is the `most likely' given the data observed, i.e., the parameter that maximizes the likelihood function of the BGM model. We assume that the data vectors are independent random variables, and thereby the log-likelihood function of $f_{\mathcal{BM}}$ can be expressed as \cite{lee2012algorithms}

\begin{equation}\label{eq:likelihood}
\begin{split}
\mathcal{L}_\mathcal{B}(\bm{\Theta}) &= \ln\prod_{n}\sum_{k}z_k^n \eta_k f_k(\bm{\xi^n})\\
&=
\sum_{n} \sum_{k}
z_k^n[\ln \eta_k+\ln f_k(\bm{\xi^n})-
\ln \int_{\bm{b_l}}^{\bm{b_u}}f_k(\bm{u})d\bm{u}]
\end{split}
\end{equation}

The expectation of $ \mathcal{L}_\mathcal{B}(\bm{\Theta}) $ can be calculated from

\begin{equation}\label{eq:expection}
\begin{split}
\mathcal{Q}_\mathcal{B}(\bm{\Theta^{(i+1)}};\bm{\Theta^{(i)}}) & =
\mathbb{E}[\mathcal{L}_\mathcal{B}(\bm{\Theta})|\bm{\xi^{1:N}};\bm{\Theta^{(i)}}]  \\
&= \sum_{n} \sum_{k} 
\langle z_k^n \rangle  \Big[ \ln \eta_k+\ln f_k(\bm{\xi^n})  \\ 
&\ \ \ \ \ \ \ \ \ \ \ \ \ \ \ \ \ -\ln \int_{\bm{b_l}}^{\bm{b_u}}f_k(\bm{u})d\bm{u} \Big]
\end{split}
\end{equation}
where the latent variable  $ \langle z_k^n \rangle:=\mathbb{P}(z_k^n=1|\bm{\xi^n}) $.

Due to the function's nonlinearity with respect to model parameter $ \bm{\Theta} $, it is extremely impossible to directly differentiate and maximize the likelihood function. To solve this problem, as shown in \cite{Lee2012EMData}, an extended EM iteration approach is employed and given by 

\begin{align}
\eta_k&=\frac{1}{N}\sum_n \langle z_k^n \rangle\\
\bm{\mu_k}&=\frac{\sum_n \langle z_k^n \rangle \bm{\xi^n}}
{\sum_n \langle z_k^n \rangle}-\bm{m_k}\\
\bm{\Sigma_k}&=\frac{\sum_n \langle z_k^n \rangle (\bm{\xi^n}-\bm{\mu_k})}
{\sum_n \langle z_k^n \rangle}+H_k
\end{align}
where
\begin{align}
\bm{m_k} &= \mathcal{M}^1(0,\Sigma_k;[\bm{b_l}-\bm{\mu_k},\bm{b_u}-\bm{\mu_k}])\\
H_k &= \Sigma_k-\mathcal{M}^2(0,\Sigma_k;[\bm{b_l}-\bm{\mu_k},\bm{b_u}-\bm{\mu_k}])
\label{eq:moment}
\end{align}
with $\mathcal{M}^1$ and $\mathcal{M}^2$ represent the first order and second order moment generated function of $f_{\mathcal{BM}}$. Update (\ref{eq:expection}) -- (\ref{eq:moment}) for each iteration and check the changed value of (\ref{eq:likelihood}). When the changed value of (\ref{eq:likelihood}) for two adjacent iterations is smaller than a predefined threshold $ \varepsilon = 1.0 \times 10^{-6} $, then stop iteration and output the optimal model parameter $ \bm{\Theta}^{\ast} $.

The component number $ K $ of the BGM model is chosen based on numerical analysis of the Bayesian Information Criterion (BIC) \cite{Box2015TimeControl}. A large value of $ K $ will increase the computational cost of training and a small value of $ K $ is unable to fully describe the underlying features of data.  As shown in Fig. \ref{fig:BIC}, the BICs for both of left and right lane departure data decrease very slowly and starts to oscillate when the number of components is greater than $ 10 $. Therefore, we chose $K=10$ in this paper by considering the computational complexity and model accuracy.

%



 
\section{Evaluation of Lane Departure Correction Systems}
An LDC system is designed to show the benefits of the proposed evaluation framework. First, we show how to regenerate lane departure events from the learned BGM model. And then, we design a controller for an LDC system based on a bicycle-vehicle model.

\subsection{Regeneration of Lane Departure Events}
A batch of lane departure events is generated from the BGM model using a simple sampling method. \textbf{Algorithm \ref{Algorithm1}}. provides a detailed procedure for regeneration of lane departure event, and discusses as follows:

\begin{enumerate}
	\item We reproduce the lane departure events that have the same statistical characteristics as the lane departure data collected from on-road experiments. The Matlab function (i.e., $ \mathtt{random} $) for generating random number was directly employed by setting the random number as $ \mathtt{N_{gen}} = 10^{5}$, see Step 2 in \textbf{Algorithm \ref{Algorithm1}};
	\item Applied (\ref{eq:boundary}) to extracting the available events generated from the learned BGM model;
	\item Reproduce lane departure events using the proposed model (\ref{eq:laterdis})--(\ref{eq:curvature}).
\end{enumerate}
By following \textbf{Algorithm \ref{Algorithm1}}, we generate a wide range of lane departure events.

\begin{algorithm}[htbp]
	\begin{algorithmic}[1]
		\State \textbf{Initialize:} Learn the BGM model and save as $ \mathtt{BGMM_{Learned}} $
		\State \textbf{Primary Samples:} Generate the random values that have the same distribution with  $ \mathtt{BGMM_{Learned}} $ by using Matlab function $ \mathtt{random} $
		
		$ \mathtt{BGMM_{Sample} = random(BGMM_{Learned}, N_{gen})} $
		\State \textbf{Check Available Values:} Extract the available vectors using the criteria (\ref{eq:boundary}), 
		
		$ \mathtt{LDE_{Variable} = MyFilter(BGMM_{Sample}, Criteria(\ref{eq:boundary}))} $
		\State \textbf{Generate Lane Departure Events:} Based on filtered variables, we can generate the lane departure events using the $ \mathtt{LaneDepModel} $ (\ref{eq:laterdis})--(\ref{eq:curvature}):
		
		$ \mathtt{LDEs = MyGenerator(LDE_{Variable}, LaneDepModel)} $
		\State \textbf{Return:} Output the reproduced lane departure events $ \mathtt{LDEs} $. 
	\end{algorithmic}
	\caption{Steps of Reproducing Lane Departure Events}
	\label{Algorithm1}
\end{algorithm}

 \begin{figure}[t]
 	\centering
 	\includegraphics[width=0.9\linewidth]{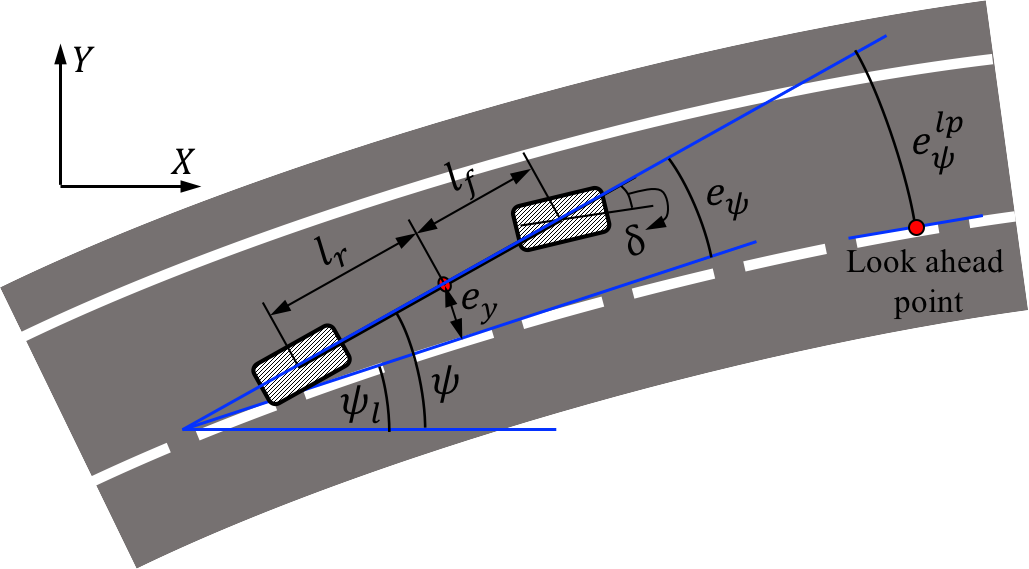}
 	\caption{Bicycle model and the relevant variables on a curve path.}
 	\label{fig:vehDyn}
 \end{figure}

\subsection{Vehicle Dynamics Model}
In this paper, a simple bicycle-vehicle model (Fig. \ref{fig:vehDyn}) is applied to validating our proposed evaluation approach. From the statistical results in Fig. \ref{fig:marginal_L} and Fig. \ref{fig:marginal_R}, we note that during the departure behavior, the curvature, curvature change, and the vehicle speed change are very small, which means the vehicle side slip angle will be very small. When vehicle speed changes slightly and the side slip value is small, a linear vehicle model can be employed. As shown in Fig. \ref{fig:vehDyn}, a simplified vehicle model is used, where two front wheels and two rear wheels are lumped together, respectively. $\psi$ is heading angle and $\psi_{l}$ is the tangent direction of the lane. We define the heading error $ e_\psi $ and offset error $ e_y $ as 

\begin{equation}
\begin{split}
e_\psi & =\psi-\psi_l \\
e_y & =y-\frac{w_v}{2}+\frac{w_{l}}{2}
\end{split}
\end{equation}
where $w_v$ and $w_{l}$ are the vehicle width and the lane width, respectively. Thus we can write the vehicle dynamic model in a state space form:

\begin{equation}
\label{eq:VehivleDynamic}
\dot{\bm{x}}(t)=\bm{A}\bm{x}(t)+\bm{B}\delta(t)+\bm{E}\dot{\psi}_l(t),
\end{equation}
where
\begin{gather*}
\resizebox{\hsize}{!}{$
	\bm{A} = 
	\begin{bmatrix}
	0 & 1 & 0 & 0\\
	0 & -\frac{2 C_{\alpha f}+2 C_{\alpha r}}{M v_x} & \frac{2 C_{\alpha f}+2 C_{\alpha r}}{M} 
	& -\frac{2 C_{\alpha f} l_f+2 C_{\alpha r}l_r}{M v_x}\\
	0 & 0 & 0 & 1\\
	0 & -\frac{2 C_{\alpha f} l_f-2 C_{\alpha r}l_r}{I_z v_x} & \frac{2 C_{\alpha f} l_f-2 C_{\alpha r}l_r}{I_z} 
	& - \frac{2 C_{\alpha f} l^2_f+2 C_{\alpha r}l^2_r}{I_z v_x}
	\end{bmatrix} $}\\
\bm{B} = 
\begin{bmatrix}
0\\
\frac{2 C_{\alpha f}}{m}\\
0\\
\frac{2 C_{\alpha f}l_f}{I_z}
\end{bmatrix}
,\qquad 
\bm{E}= 
\begin{bmatrix}
0\\
-\frac{2 C_{\alpha f}l_f-2 C_{\alpha r}l_r}{m v_x}-v_x\\
0\\
-\frac{2 C_{\alpha f}l^2_f+2 C_{\alpha r} l^2_r}{I_z v_x}
\end{bmatrix}
\end{gather*}
where $\bm{x}(t)=[e_y,\dot{e}_y,e_\psi ,\dot{e}_\psi]^{\top}\in \mathbb{R}^{4\times 1},\ \bm{A}\in \mathbb{R}^{4\times 4},\ \bm{B}\in \mathbb{R}^{4\times 1},\ \bm{E}\in \mathbb{R}^{4\times 1}$, $C_{\alpha f}$ and $C_{\alpha r}$  are the tire stiffness, $l_f$ and $l_r$ are the longitudinal distance from center of gravity to the frontal axle and rear axle, respectively, $m$ is the total mass, $I_z$  is the inertia of $z$ axis,  and $\delta$, as the model input,  is the steering angle of the front wheel.



\subsection{Controller Design}
The main goal of designing a controller is to validate the efficiency of our proposed stochastic driver model. Thus, we design an aim point controller \cite{falcone2011predictive} for the steering system to help driver pull the vehicle back to lane center. We define the preview orientation error $ e_\psi^{lp} $ (Fig. \ref{fig:vehDyn}) as

\begin{equation}\label{eq:25}
e_\psi^{lp} = \psi-\psi_l^{lp}
\end{equation}
and assign

\begin{equation}\label{eq:26}
\Delta\psi_l=\psi_l-\psi_l^{lp}
\end{equation}

Substituting (\ref{eq:26}) to (\ref{eq:25}), we have

\begin{equation}
e_\psi^{lp} = \psi-(\psi_l-\Delta\psi_l) = e_\psi+\Delta\psi_l
\end{equation}

Considering the control law

\begin{equation}
\delta = K_y e_y+K_\psi e_\psi^{lp}
\end{equation}
 and letting $\bm{F} = [K_y,0,K_\psi,0]$ and $G=K_\psi$, we have

\begin{equation}
\label{eq:ctrl}
\begin{split} 
\delta & = K_y e_y+K_\psi e_\psi+K_\psi\Delta\psi_l\\
& = \bm{F}\bm{x}+G\Delta\psi_l
\end{split}
\end{equation}

The LDC system kicks in when a predefined threshold is valid. In this paper, we activate the designed controller when  $y<y^{R}_{s}<0$ for the right lane departure or $y>y^{L}_{s}>0$ for the left lane departure is valid, where $ y^{R}_{s} $ and $ y^{L}_{s} $ are both the predefined thresholds. We set $ |y^{L}_{s}| = |y^{R}_{s}| = y_{s} = 0.2 $ m. After the vehicle is pulled back to around the center of the driving lane, the controller will stop work. Based on the discussion above, we rewrite (\ref{eq:VehivleDynamic}) as follows

\begin{equation}\label{eq:VehivleDynamic2}
\dot{\bm{x}}(t)=\bm{A_c}\bm{x}(t)+\bm{B_c} \bm{\varPsi}(t),
\end{equation}
where $\bm{A_c}=\bm{A}+\bm{B}\bm{F}$, $\bm{B_c}=[\bm{E},\bm{B}G]$, $ \bm{\varPsi}(t)=[\dot{\psi}_l(t)$, $\Delta\psi_l(t)]^{\top}$. Substitute  (\ref{eq:ctrl}) into (\ref{eq:VehivleDynamic2}), we obtain the close loop form

\begin{equation}
\dot{\psi}_l(t)=v_x(t_s)c(t)=\frac{v_x(t_s) \Delta c}{T}t+v_x(t_s)c_0
\end{equation}

\begin{equation}
\Delta\psi_l(t)=\int_{t} ^{t+T_{lp}} \dot{\psi}_{l}(t)  dt
=A_{\Delta\psi_l} t+B_{\Delta\psi_l}
\end{equation}
where $A_{\Delta\psi_l} = \frac{\Delta c \cdot T_{lp} \cdot v_x(t_s)}{T}$ and $B_{\Delta\psi_l}=\frac{\Delta c\cdot T_{lp}^2 \cdot v_x(t_s)}{2 T}+v_x(t_s)\cdot c_0\cdot T_{lp}$. The initial condition of the close loop control is $e_y(t_s)=y(t_s)+(w_l-w_v)/2$ and $e_\psi(t_s) = \arctan(\frac{v_y(t)}{v_x(t)})$, where $v_y(t_s) = \frac{d e_y(t)}{dt}|_{t = t_s} $, $ t_s $ is the starting time of lane departure happening. Table \ref{tab:simpara} gives the parameters used in the simulation.

\begin{table}[t]
	\caption{Simulation Parameters}
	\label{tab:simpara}
	\centering
	\begin{tabular}{c c c|c c c}
		\hline\hline
		Var & Unit & Value & Var & Unit & Value\\
		\hline
		$C_{\alpha f}$ &N/rad	&80,000	&$l_{r}$	&m			&1.47	\\
		$C_{\alpha r}$	&N/rad	&80,000	&$I_z$		&$kgm^2$	&3,344	\\
		$l_{f}$			&m		&1.43	&M			&kg			&1,000	\\
		$T_lp$			&s		&2		&$D_y$		&m			&0.5	\\
		$T_s$		&s		&0.1 &$w_l$		&m		& 3.6 \\
		$K_y$		&rad/m	&-0.005 & $ w_v $ & m & 1.9  \\
		$K_\psi$	&rad/rad&-0.2 & $ y_s $ & m & 0.2\\
		\hline
		\hline
	\end{tabular}
\end{table}

\section{ Analysis and Discussions for Simulation Results}
We analyze our proposed evaluation framework from two aspects: (1) statistical analysis of reproduced lane departure events and (2) evaluation of the controller using the reproduced lane departure events.

\begin{figure}[b]
	\centering
	\includegraphics[width = 0.48\textwidth]{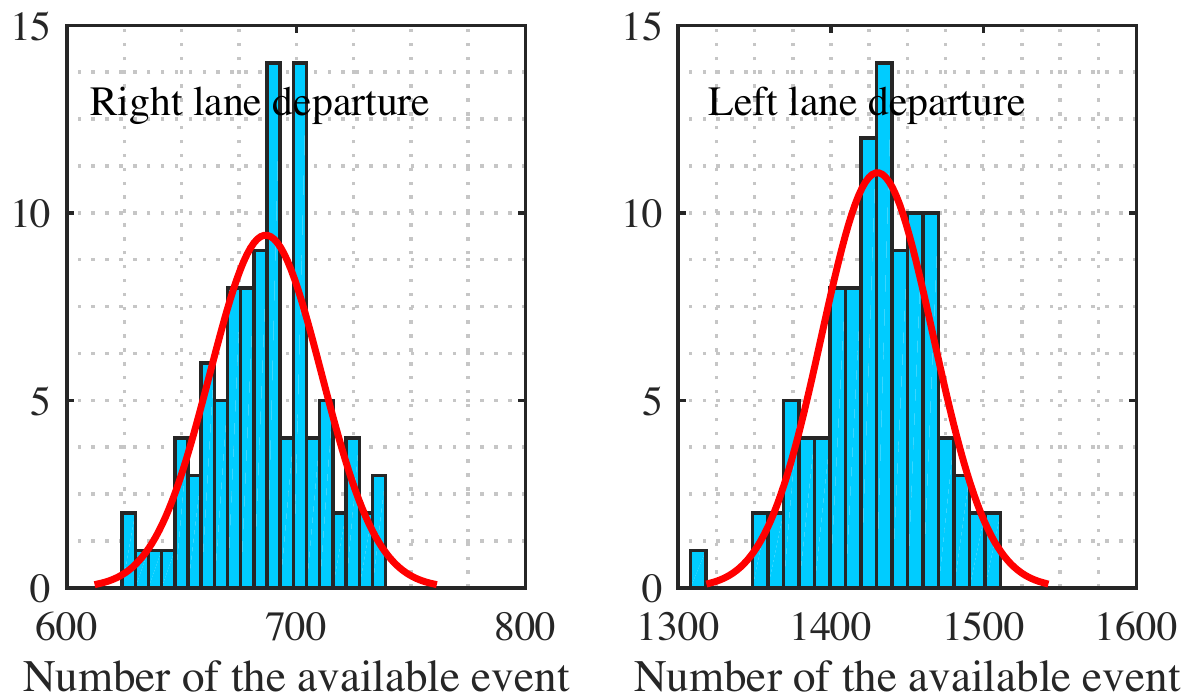}
	\caption{Statistical results for available lane departure events generated from $ 10^{5} $ events.}
	\label{fig:eval_LDEs}
\end{figure}

\begin{figure}[t]
	\centering
	\includegraphics[width = 0.48\textwidth]{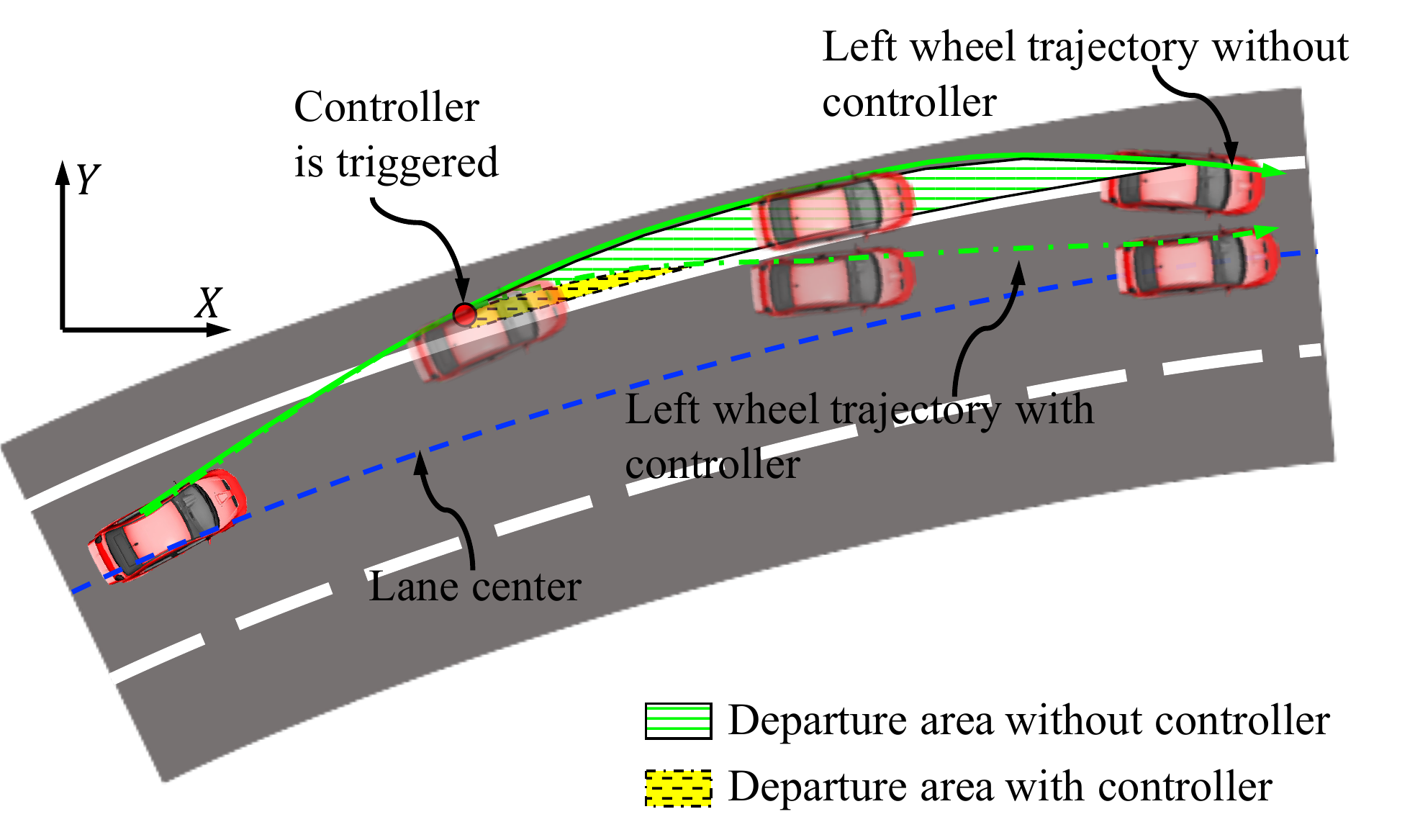}
	\caption{Illustrations of vehicle trajectories with and without the designed controller.}
	\label{Illus_ex}
\end{figure}

\begin{figure}[t]
	\centering
	\begin{subfigure}[t]{0.48 \textwidth}
		\centering
		\includegraphics[width = 1\textwidth]{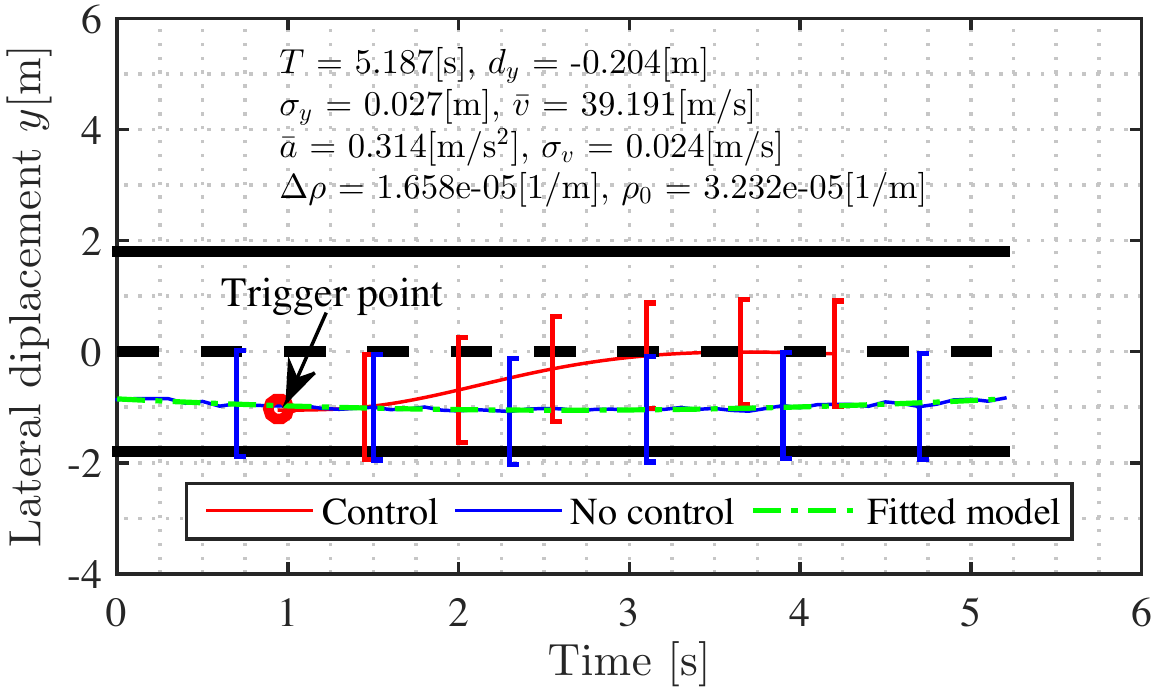}
		\caption{Right departure cases}
	\end{subfigure}
	\begin{subfigure}[t]{0.48 \textwidth}
		\centering
		\includegraphics[width = 1\textwidth]{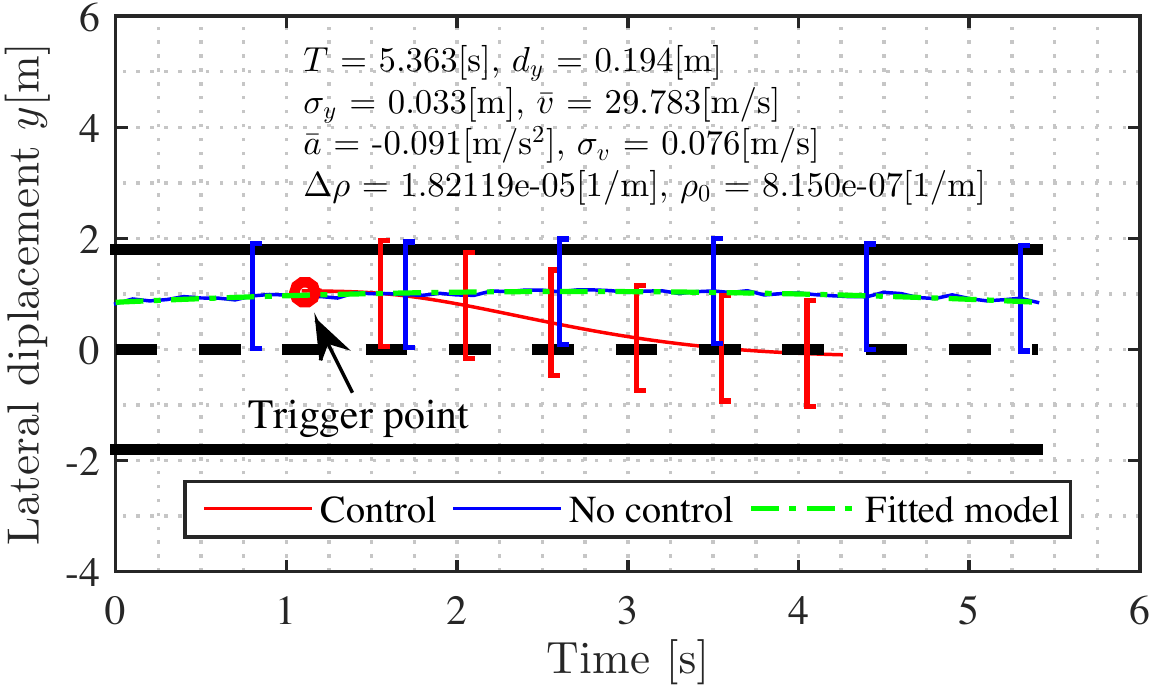}
		\caption{Left departure cases}
	\end{subfigure}
	\caption{The comparison results of the lane departure events with and without controller.}
	\label{Result_Ex}
\end{figure}

\begin{figure*}[t]
	\begin{subfigure}[t]{0.68 \textwidth}
		\centering
		\includegraphics[scale = 0.7]{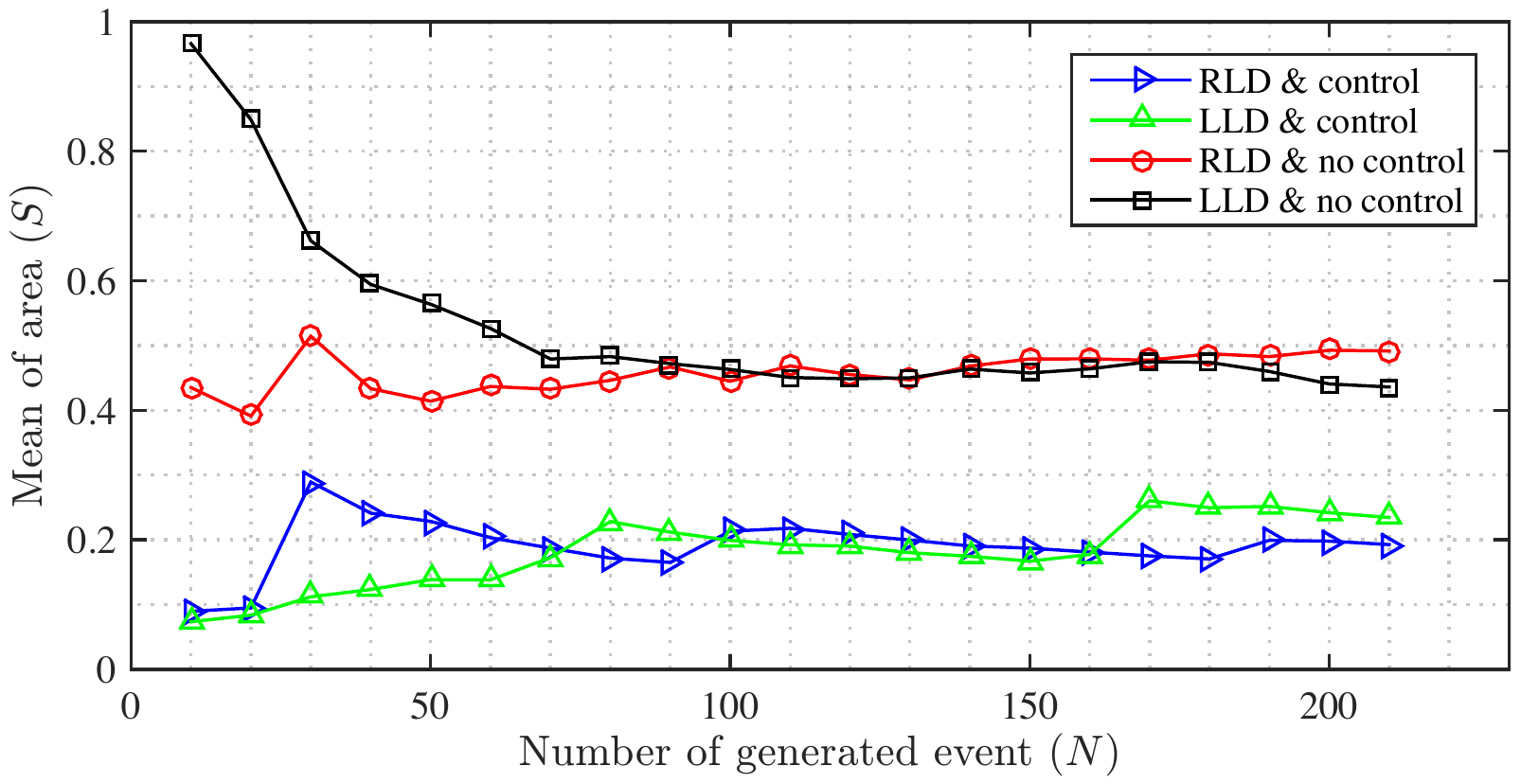}
	\end{subfigure}
	\begin{subfigure}[t]{0.28 \textwidth}
		\centering
		\includegraphics[scale = 0.7]{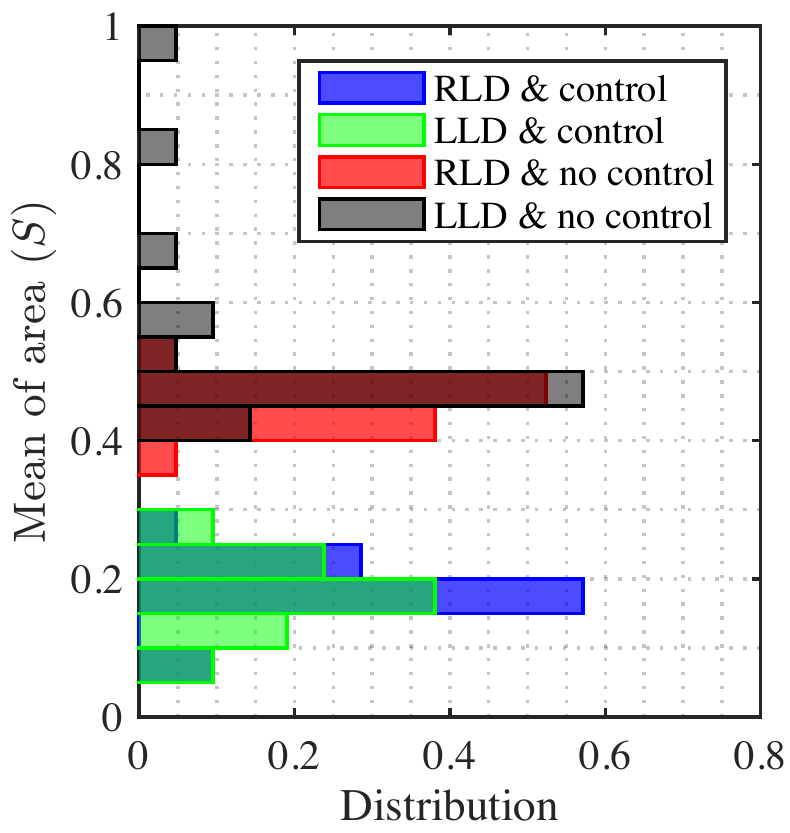}
	\end{subfigure}
	\caption{The statistical results for different amounts of lane departure event, including left lane departure (LLD) event and right lane departure (RLD) event.}
	\label{Result_stat}
\end{figure*}

\subsection{Reproduction Efficiency for Lane Departure Events}
We evaluated the efficiency to reproduce lane departure events. Fig. \ref{fig:eval_LDEs} shows the statistical results of available events in all lane departure events when we repeatedly run the lane departure model for 100 times. We find that about 7\% and 14.5\% of $ 10^{5} $ reproduced events satisfy the condition (\ref{eq:boundary}) for right and left lane departure behaviors, respectively. This approach generates lane departure behaviors at a relatively low proportion, but it computes very fast and takes totally less 2 s to generate $ 10^{5} $ lane departure events using MatLab in the laptop with a processor of 2.5 GHz Intel Core i7. Our proposed dimension reduction approach in Section II significantly reduces the computational time.
%
%

\subsection{Controller Evaluation}

Fig. \ref{Illus_ex} illustrates the left lane departure events (green solid line) generated from the stochastic model and the trajectories (red dot line) of vehicle's left front wheel with the designed controller. The red point represents the trigger point where the control condition is valid, i.e., $ y <-0.2  $ m for right departure behavior or $ y>0.2 $ m for left departure behavior. 

Fig. \ref{Result_Ex} provides two examples of simulation results for right and left departure behaviors with/without the controller. The black solid line represents the lane boundary and the black dash line is the center of the driving lane. The blue line is the lane departure trajectory generated from the stochastic lane departure model. The green dash line is the fitted model using (\ref{eq:laterdis})--(\ref{eq:curvature}), and the red line is the vehicle trajectory with the designed controller. In Fig. \ref{Result_Ex}, when the condition of vehicle lateral displacement is valid (i.e., $ y_L>y_s $ or $ y_R<-y_s $, denoted by red circles in Fig. \ref{Result_Ex}),  the controller is triggered, thereby assisting the driver in steering the vehicle back to the center of the driving lane. It is obvious that the proposed stochastic lane departure model can reproduce the lane departure events similar to these in the real world, and then can be used to evaluate a controller performance for LDC systems at a low cost. 

To further show the benefits of our proposed evaluation framework, the area between the trajectory of vehicle's left(right) wheel for left(right) lane departure behaviors and lane boundary for each departure event is defined and computed by 
%
%
%
%

\begin{equation}\label{eq:areas}
S = \int_{t_{start}}^{t_{end}}   |{y}(t)|  dt
\end{equation}
where $ t_{start} $ and $ t_{end} $ are the starting and ending time of the departure behaviors, respectively, with a smaller value of $ S $ indicating that the vehicle is tracking the lane center better. We use (\ref{eq:areas}) to evaluate the controller performance for left and right departure behaviors. The evaluation metric (\ref{eq:areas}) may not be the most accurate approach to measure the effectiveness of the LDC systems, which may need to consider a variety of aspects such as yaw stability, driving speed, road conditions. However, since the proposed method provides the kinematic information of the vehicle, it can be easily extended to include the aforementioned factors once the extra data is available.

\begin{figure}[t]
	\centering
	\includegraphics[width = 0.43\textwidth]{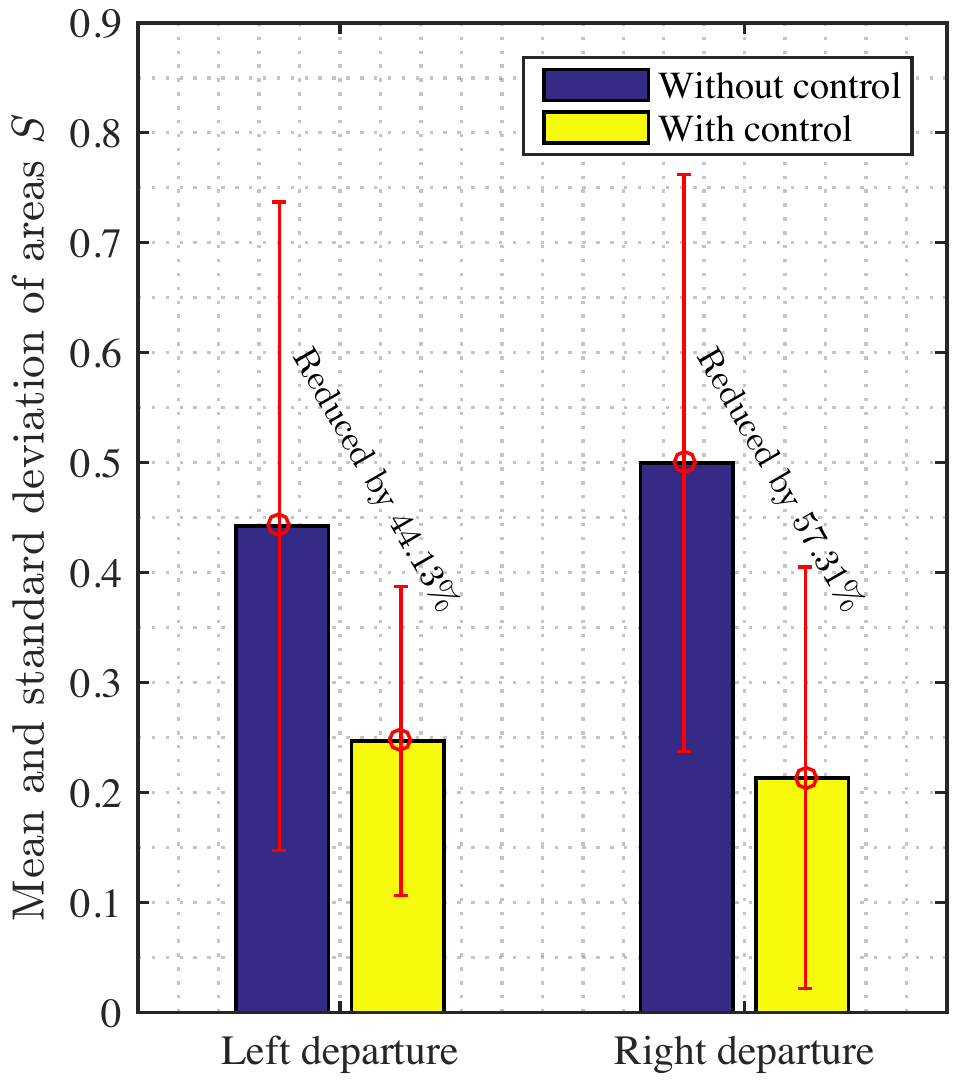}
	\caption{The mean and standard deviations of departure covering areas for 200 left and right departure events, respectively.}
	\label{Result_area}
\end{figure}

Fig. \ref{Result_stat} shows the statistical results of vehicles with/without the designed controller with respect to different numbers of lane departure events, including right lane departure and left lane departure cases. The average values of the area ($ S $) for both right and left lane departure events with or without controller are convergent. Drivers usually prefer to take a left lane departure when driving as the mean of the area ($ S $) for right lane departure behaviors without controller (red line) is smaller than that for left lane departure behaviors (black line).

Fig. \ref{Result_area} provides the statistical simulation results of 200 left lane departure events and 200 right lane departure events that are both randomly produced from the stochastic lane departure model. The bar represents the mean value and the red vertical line represents the standard deviation of $ S $. It can been seen that our proposed method can cover various lane departure events and that the controller can help drivers steer vehicles back to the lane center when drivers depart away from the lane center. Vehicles with LDC systems can reduce the departure areas $ S $ by $ 44.13 \%$ and $ 57.31 \%$ for right and left departure behaviors, respectively. 

\subsection{Further Discussions and Future Work}
The proposed lane departure model can capture the uncertainties of human driving and improve the computational efficiency. In this paper, we applied the stochastic driver model to evaluating LDC systems. When the departure behavior occurs, the controller is activated and takes the place of human driver's control. In addition to this application, the proposed driver model can also be used to
\begin{enumerate}
	\item assess the reliability and acceptability of LDW systems with different designed waring strategies at a low cost;
	\item investigate into the influence of different ADASs on human drivers' behaviors such as decision making and reactions;
	\item develop a driving simulator, enabling the lane departure behaviors of other vehicles to be highly similar to those in the real world.
\end{enumerate}
Therefore, the work presented in this paper has the potential to be widely used in the evaluation of autonomous vehicles.

\section{Conclusions}
In this paper, we propose a framework for evaluating lane departure correction (LDC) systems based on a stochastic driver model using a huge amount of naturalistic driving data. A stochastic driver model that can reproduce various lane departure events as similar to what drivers will do in real-life driving situations is developed by considering the uncertainties of human driving. In the stochastic driver model, in order to improve the computational efficiency, we propose a dimension reduction method using a polynomial function with the addition of stochastic terms characterizing uncertainties in human driving. And then, we apply the bounded Gaussian mixture model to fitting the data with incorporating the bounded characteristics of physical variables. Last, to show the benefits of our proposed evaluation framework, a controller is designed and applied to a bicycle-vehicle model. The simulation results show that the controller can assist a driver in steering vehicles back to the center of the driving lane when departure behavior occurs. This supports that the proposed stochastic driver model is an efficient way to evaluate LDC systems at a low cost.

The proposed approach can be generalized to evaluate other ADASs such as the autonomous emergency braking system or the lane change assistance system. It also has the potential to evaluate autonomous vehicles by simulating the ambient stochastic driving environment.


%

%

%

 						


\ifCLASSOPTIONcaptionsoff
  \newpage
\fi



\bibliographystyle{IEEEtran}
\bibliography{Mendeley_Lane_Departure}
%
%
%

%

\begin{IEEEbiography}[{\includegraphics[width=1in,height=1.25in,clip,keepaspectratio]{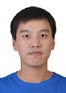}}]{Wenshuo Wang}
received his B.S. in Transportation Engineering from Shandong University of Technology, Shandong, China, in 2012. He is a Ph.D. candidate in Mechanical Engineering at the Beijing
Institute of Technology (BIT). He is a currently visiting scholar in the School of Mechanical Engineering, University of California at Berkeley (UCB). He conducts research under the supervision of Prof. Junqiang Xi (BIT) and Prof. Karl Hedrick in the Vehicle Dynamics \& Control Lab, University of California at Berkeley. His research interests include vehicle dynamics control, adaptive control, driver model, human-vehicle interaction, recognition and application of human driving characteristics. His work focuses on modeling and recognizing drivers behavior, making intelligent control systems between human drivers and vehicles.
\end{IEEEbiography}

\begin{IEEEbiography}[{\includegraphics[width=1in,height=1.25in,clip,keepaspectratio]{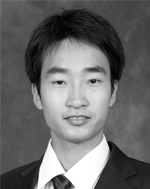}}]{Ding Zhao} received his Ph.D. degree in 2016 from the University of Michigan, Ann Arbor. He is currently a Research Fellow in the University of Michigan Transportation Research Institute. His research interests include the evaluation of connected and automated vehicles, vehicle dynamic control, driver behaviors modeling, and big data analysis.
\end{IEEEbiography}
%
%

%
%




\end{document}